# Long range fading free phase-sensitive reflectometry based on multi-frequency NLFM pulse

JINGDONG ZHANG,[1][†] HUA ZHENG[1], TAO ZHU,[1][*] GUOLU YIN,[1] MIN LIU,[1] DINGRONG QU,[2] FENG QIU,[2] AND XIANBING HUANG[2]

[1]Key Laboratory of Optoelectronic Technology & Systems (Ministry of Education), Chongqing University, Chongqing 400044, China
[2]State Key Laboratory of Safety and Control for Chemicals, SINOPEC Research Institute of Safety Engineering, Qingdao 266000, China

*Corresponding author: zhutao@cqu.edu.cn    †: zjd@cqu.edu.cn

A long range phase-sensitive optical time domain reflectometer (φ-OTDR) with a multi-frequency non-linear frequency modulation (NLFM) optical pulse is proposed in this Letter. To boost the pulse energy while suppressing the optical nonlinear effects, the distortion of the amplified pulse is rectified, and a three-tone pulse is used. Combining with the NLFM technic which provides 42.7 dB side lobe suppression ration (SLSR), these two approaches guarantee that a sensing distance of 80 km is achieved in the experiment with 2.5 m spatial resolution, 49.6 dB dynamic range, and 45 dB phase signal-to-noise ratio (SNR). To the best of our knowledge, this is the first time that a phase-demodulated φ-OTDR over such a long sensing range has been reported with un-pumped sensing fiber.

Phase-sensitive optical time domain reflectometry has attracted considerable attention in the last decades due to its capability for the distributed monitoring of acoustic or vibration induced strain over long fiber distance, which can potentially be applied in perimeter protection, mine micro-seismic monitoring, and pipeline surveillance [1-5]. At present, many significant researches have been done to enhance the sensing performance of φ-OTDR. Specifically, ultra-long φ-OTDR with 175 km sensing range is realized by distributed amplification [1]; the spatial resolution is improved to less than 1 meter with frequency-swept pulse [2,3] or PSK pulse [4]; the frequency response range is increased with frequency division multiplexing [5] or non-uniform sampling [6]; the phase signal induced by the perturbation can be extracted by digital coherent [7] or optical I/Q detection [8]; the fading phenomena can be partially eliminated by polarization-maintain configuration [9] or using multi-frequency source [10]; and so on.

As for long range φ-OTDR sensing, the interrogating configurations developed so far can be classified in terms of whether they adopt amplification in the sensing fiber. Though distributed amplifications, including Raman amplification [11], Brillouin amplification [12], and their combination [13], have been validated to compensate the fiber loss and thus extend the sensing distance to over 100 km, these configurations require the pump light be injected into the both ends of sensing fiber, which reduces the degree of freedom in embedding the fiber into structures. The un-pumped schemes are characterized by single ended injection, but their sensing distances are limited to within 50 km [14,15]. However, frequency modulation technique is promising for sensing range improvement, because it increases the pulse length without deteriorating the spatial resolution [16]. Furthermore, the phase signal, which is linear response to the external vibration, can also be extracted during the signal demodulation [17-18].

In this work, a phase-sensitive reflectometry based on multi-frequency NLFM pulse is presented, in which a sensing range of 80 km and spatial resolution of 2.5 m are proved to be feasible simuitaneously without distributed or remote amplification. This system is realized from three aspects. First, the crosstalk of the adjacent backscattering points is suppressed using the NLFM pulse whose side lobe suppression ration (SLSR) is 42.7 dB. Next, the distorted laser pulse after amplified by erbium-doped fiber amplifier (EDFA) is corrected in a iterative pre-distortion approach, and thus the non-liner effects like self-phase modulation (SPM) are suppressed. Finally, by mudulating the laser to a multi-frequency sourse with electro-optic modulator (EOM), the coherent fading effect as well as Brillouin scattering is greatly reduced. A verification expreiment is implemented using a three-tone NLMF pulse with 60 MHz sweeping ranges, and up to 600 Hz phase signals induced by vibration are quntified with 45 dB SNR.

The proposed system, as well as reported in [2-4,16-18], is derived from the matched filter theory which is central to signal processing in fields such as radar, communication, and sonar. The matched filter is the optimal linear filter for the signal with white noise, and it provids criteria for filter design with high SNR. Denoting $h(t)$ as the impulse response of the matched filter, the output $y(t)$ of the filter for an input signal $s_i(t)$ with white noise $n_i(t)$ can be written as $y(t)=s_o(t)+n_o(t)$ accroding to linear system superposition theorem, in which $s_i(t)$, $n_i(t)$ are the complex envelopes of the input waveform and white noise. Specifically, if the double side band power spectrum density of the white noise is $N_0$, then the SNR of the signal at time $t_0$ is

$$SNR = \frac{P_s}{P_n} = \frac{s_o(t_0)s_o^*(t_0)}{E\left[n_o(t)n_o^*(t)\right]} \leq \frac{E_s}{N_0}, \quad (1)$$

where $P_s$ is the power of $s_o(t)$ at time $t_0$, $P_n$ is the average power of $n_o(t)$ and $E_s$ is the energy of $s_i(t)$ [19]. The left side of Eq. (1) equals the last term of Eq. (1) when $h(t)=As_i^*(t_0-t)$ with $A$ is a real constant, and then the SNR gets its maxmium value. Thus, $s_i^*(-t)$ is regarded as the matched filter of the system. Equation (1) also reveals a remarkable result that the achievable SNR depends only on the energy of the input waveform, but not on other details such as modulation type, which provides a criteria of the probe waveform designing for φ-OTDR with long interrogation range.

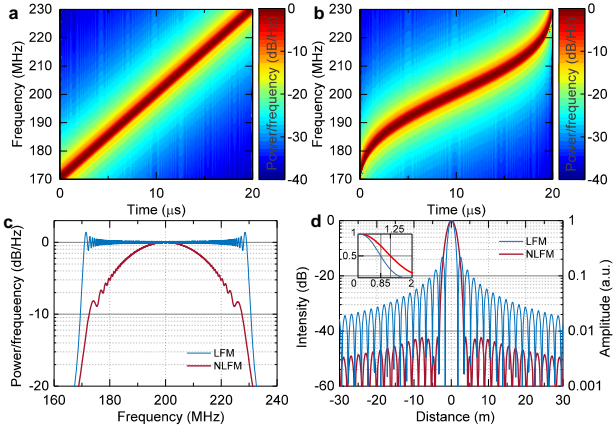

**Fig. 1.** Time-frequency spectra of LFM (a) and NLFM (b). (c) Relative frequency spectra of LFM and NLFM. (d) Compressed pulses.

Based on the matched filter theory and the criteria it established, frequency modulated pulse serves as a significant substitute for the typical single frequency prob pulse in φ-OTDR sensing, and provides another substantial benefit of pulse compression [16] which enables a high spatial resolution with a large time-width pulse. With non-linear phase profile, the chirped pulse possesses a large frequency bandwidth which is irrelevant to its time-width, and thereby a higher spatial resolution after matched filtering. Specifically, linear frequency modulation (LFM) pulse with quadratic phase shape is one of the most feasible modulation types. Denoting

$$E(t) = \text{rect}(\frac{t}{T})\exp(j2\pi f_c t + \frac{j\pi Bt^2}{T}) \quad (2)$$

as the complex envelope of the LFM probe light pulse, the spatial resolution of the system after matched filtering is $R=c/(2n_{eff}B)$ [17], in which $T$ is the time-width of the pulse, $f_c$ is the central frequency of the laser, $B$ is the sweeping frequency band, $c$ is the light speed in vacuum, $n_{eff}$ is the effective refractive index of fiber. The time-frequency spectrum of LFM is shown in Fig. 1(a), and the compressed pulse is plotted in Fig. 1(d). With $B$=60 MHz and $f_c$=200 MHz, the LFM reaches a spatial resolution of 1.7 m. However, the SLSR is only 13.2 dB, which will lead to strong signal crosstalk of the adjacent backscattering points [18]. To enhance the SLSR, typically, a window function is applied to the amplitude of pulse signal, or to the LFM matched filter during demodulation, which will cause a decrease of signal energy or a mismatching loss [19]. As shown in Fig. 1(c), the NLFM pulse, applied Hamming window function in frequency domain, is adopted in this paper. According to the stationary phase method [19], the instantaneous frequency of NLFM can be obtain as shown in Fig. 1(b). Without reducing the pulse energy or breaking the matched filter condition, the SLSR is enhanced to 42.7 dB, as displayed in Fig. 1(d).

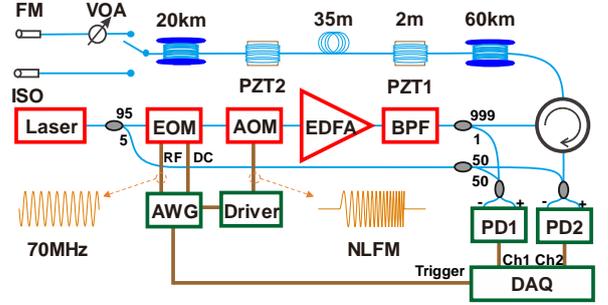

**Fig. 2.** Experiment setup.

The experiment setup of a long-range NLFM φ-OTDR characterized by fading free is shown in Fig. 2. The narrow linewidth laser (NKT Laser, E15, @1550.12 nm) is modulated to a multi-tone optical source by the EOM (Photline, MXER-LN-10) driven by the arbitrary waveform generator (AWG, Tektronix, AWG5012C). The RF (radio frequency) port of the EOM is driven by a 70 MHz single frequency signal with 1 Vpp amplitude. By sweeping the DC (direct current) bias, the power of each tone can be regulated as shown in Fig. 3(a), in which the $0_{th}$ tone is the optical carrier at 1550.12 nm, while $\pm 1_{th}$ and $\pm 2_{th}$ tones are $\pm 70$ and $\pm 140$ MHz away from the carrier in frequency domain. When the DC bias is ~4.8 V, the $0_{th}$ and $\pm 1_{th}$ tones reach the same power, and the optical output of the EOM is a three-tone source with 49.5 dB harmonic suppression ratio. Subsequently, this multi-frequency source is modulated to NLFM pulse by the acoustic-optic modulator (AOM, Gooch & Housege). The heterodyne signal between the AOM output and the 1550.12 nm carrier is presented in Fig. 3(b), in which the original pulse centered at 200 MHz is duplicated at 130 and 270 MHz. This multi-frequency NLFM optical pulse, severing as the light probe, guarantees that the proposed system have features of pulse compression and coherent fading free.

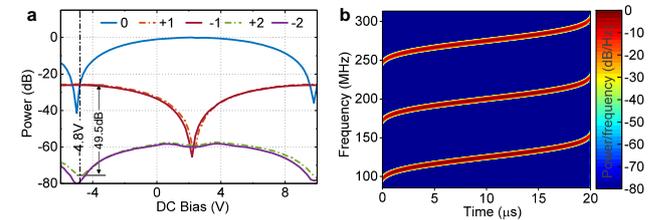

**Fig. 3.** (a) Different tone powers with DC bias. (b) Multi-frequency NLFM pulse.

As revealed by Eq. (1), the SNR of the system is proportional to the energy of the laser pulse. Hence, the pulse energy is improved, and the time width of the pulse is set to 20 us which is 800 times larger than that in conventional φ-OTDR system with 2.5 m spatial resolution. To improve the pulse power and compensate the losses of EOM and AOM, the pulse is amplified by the EDFA (Beogold, PEDFA). However, due to the transient effect of the EDFA [20], the pulse shape is distorted as shown in Fig. 4(a) (blue line marked as $m$=1). In order to study the impact of pulse distorted on the system, a single tone NLFM pulse (Fig. 1(b)) is adopted, and a Faraday mirror (FM) is set at the end of the 80 km sensing fiber. The amplified NLFM pulse is filtered by a band-pass filter (BPF, 10GHz), and injected into the sensing fiber through the optical circulator. Adjusted by the variable optical attenuator (VOA), the reflection pulse with suitable power (45dB larger than the backscattering light power at 80 km) is heterodyned with the carrier, detected by PD2 (photodetector, Thorlabs, PD480C) and then sampled by the data acquisition card (DAQ, Gage, CSE24G8) working at 1 GSa/s. The compressed pulse of the FM reflection is shown in Fig. 4(c). Compared with the theoretical line marked as green dots, the compressed pulse of the end reflection is shifted in distance and broadened in width. This is contributed by the optical nonlinear phenomena stimulated by the high power distorted pulse [21]. Firstly, the distorted pulse with a power gradient within the pulse width is shifted by SPM when passing through the sensing fiber. The distance dependent frequency shift of the pulse is unstable and will cause a positioning error for the ambiguity relationship between Doppler shift and range measurement error [19]. Furthermore, the distorted pulse with high power peak triggers stimulated Brillouin scattering which also exacerbates the mismatch of matched filter condition, and broadens the width of the compressed pulse.

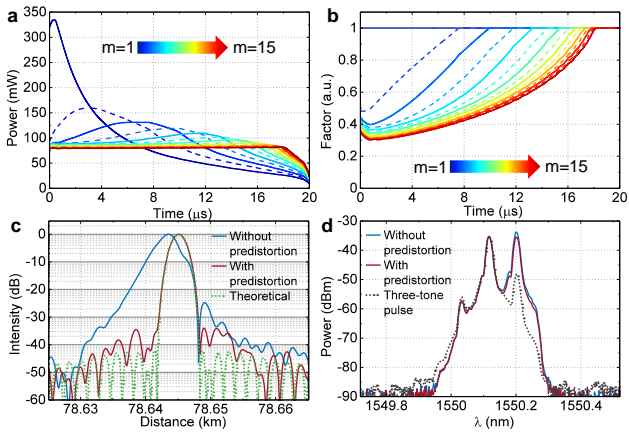

**Fig. 4.** (a) Envelopes of pulse intensities in iteration. (b) Pulse envelope factors in iteration. (c) Compression pulses of FM reflection. (d) Backscattering optical spectra.

In order to reduce the impacts of pulse distorted on pulse compression and extend the sensing range, the NLFM pulse is pre-distorted in AWG by multiplying an envelope factor onto the instantaneous amplitude of NLFM. This is achieved in a closed-loop iteration as illustrated in Fig. 4(a) and (b). By referring the different envelopes of pulse intensities in iteration, diverse envelope factors are used. After 15 times of iterations, the slope of the amplified pulse turns into zero, and the compressed pulse of the PM reflection is shown in Fig. 4(c). The positioning error and the pulse shape broadening are eliminated. The backscattering spectrum of the sensing fiber is also measured after connecting the fiber end to the optical isolator (ISO). As shown in Fig. 4(d), with same energy of 1.7 μJ, the pulses with and without predistortion shear the same Rayleigh backscattering peak power, but the Brillouin backscattering peak power of pulse with predistortion is 2 dB lower than that of pulse without predistortion. It reveals that the Brillouin backscattering is slightly suppressed by predistortion pulse. It is noted that the predistortion three-tone pulse, with the same pulse energy, will further suppress the Brillouin backscattering as shown in Fig.4(d). This is because the Brillouin gain that multi-frequency pulse provides is smaller than that of single-frequency pulse [21].

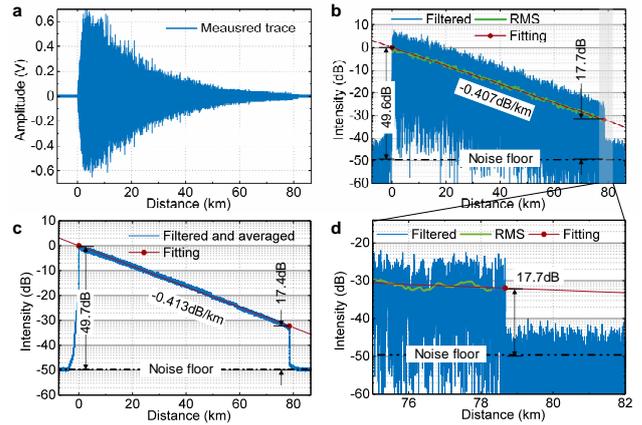

**Fig. 5.** (a) Single measured trace. (b) Single trace after matched filtering. (c) Filtered and averaged trace. (d) Enlarge figure of the end section.

Obtained the amplified pulse with boosted energy and optimum shape, the dynamic range of the system is illustrated in what follows. Figure 5(a) plots a single backscattering trace of single-tone NLFM pulse on 80 km sensing range. This trace is filtered by the matched filter and the compressed pulse is shown in Fig. 5(b) and (d). By fitting the moving root-mean-square (RMS, with 20 m moving window) of the compressed pulse, the backscattering signal is found to be reduced exponentially of ~0.4 dB/km. This corresponds to the double of the attenuation coefficient of the used fiber (0.2 dB/km). As it can be observed in Fig. 5(b) and (d), the SNRs of the front and the rear sensing end are 49.6 and 17.7 dB, in which the noise floor is defined by the RMS of the noise (containing thermal noise of PD and amplified spontaneous emission noise introduced by EDFA). By sweeping the wavelength and scrambling the polarization states of the NLFM pulse, the average of 500 backscattering traces is presented in Fig. 5(c), in which the SNRs and attenuation coefficient are in good agreement with those of single trace in Fig. 5(b) and (d).

Next, the feasibility of the presented system for vibration sensing over 80 km is demonstrated. As shown in Fig. 2, two fiber sections with length of 2 m are strapped around two piezoelectric transducers (PZT) in the 60 km of the sensing fiber. The PZT1 and PZT2 are fed with sinc function and sinusoidal signal, respectively. Three-tone NLFM pulses are injected into the fiber with a repetition rate of 1.24 kHz, allowing us to measure up to 620 Hz vibrations. The amplified pulses are also split and heterodyned with the carrier, and then detected by PD1, by which the powers of the three tones can be measured and adjusted dynamically. The heterodyned signal of backscattering light is detected by PD2 and digitized by DAQ. The signal from PD2 possesses three peaks centered at 130, 200 and 270

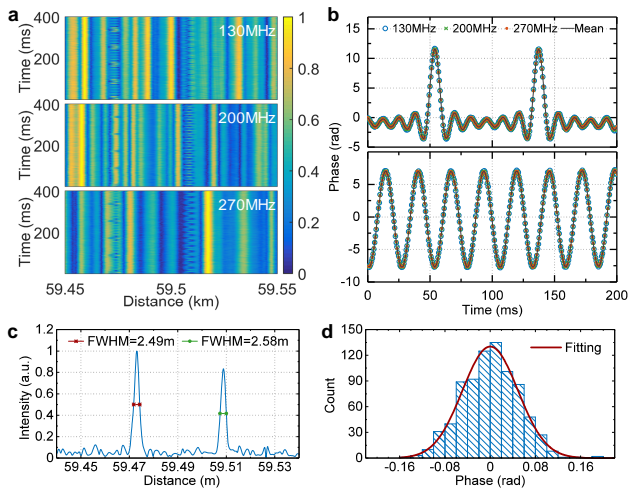

**Fig. 6.** (a) Amplitude traces of different frequencies (b) Demodulated phase signals of two vibration points. Top: PZT1, Bottom: PZT2. (c) Averaged standard deviation of the trace intensities. (d) Phase error histogram at PZT2.

MHz, respectively. Demodulated by their corresponding matched filters, the amplitude traces of the compressed signals are shown in Fig. 6(a), in which the vibration points can be located clearly. The interference fading free feature of the system can also be verified, as the fading points will not take place at same position of the three subfigures simultaneously. As shown in Fig. 6(b), the phase signals from different frequency bands are demodulated [2] and in good agreement with each other. The average signals of the phase signals are shown as gray lines in Fig. 6(b). The phase errors of average signal of PZT2 is extracted, and its histogram is presented in Fig. 6(d). Subjecting to normal distribution roughly, the phase errors have a standard deviation of 0.049 rad, which is consistent with the SNR (25.9 dB) of the backscattering trace intensity in 60 km [18]. To demonstrate the spatial resolution of the system, the strapped ranges of PZTs are shortened to 0.5 m. With vibration signals applied to the PZTs, the averaged standard deviation of the trace intensities is shows in Fig. 6(c). The spatial resolution is about 2.5 m, which is consistent with the theoretical estimate in Fig. 1(d) and 4(c).

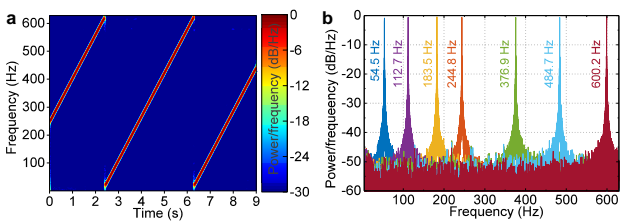

**Fig. 7.** (a) Time-frequency spectrum of frequency modulated signal. (b) Frequency spetra of different frequency signals.

Finally, a frequency-modulated (FM) signal from 10 Hz to 610 Hz is applied to PZT1, and the time-frequency spectrum of averaged phase signal is shown in Fig. 7(a). Then, diverse sinusoidal signals are applied to PZT2. The corresponding frequency spectra are well measured without harmonic components, which indicates the proposed system allows us to detect and quantify the perturbation induced phase changes with up to 600 Hz frequency range and exceeding 45 dB SNR.

In summary, a multi-frequency NLFM pulse based φ-OTDR is proposed. The probe pulse energy of the system is increased with the optical non-linear effects be suppressed by pridistortion multi-frequency technics, and hence the sensing distance of φ-OTDR is enhanced. Without pump amplification in sensing fiber, sensing range of 80 km, spatial resolution of 2.5 m, and fading free feature of the system are verified experimentally. The proposed method, providing long sensing range with single-ended sensing structure, is promising in practical vibration sensing applications.

**Funding.** Ministry of science and technology (2016YFC0801202); National Natural Science Foundation of China (61635004, 61475029, 61775023, 61405020); Science fund for distinguished young scholars of Chongqing (CSTC2014JCYJJQ40002).